\documentclass[12pt]{article}
\usepackage{amsmath,amsfonts,amssymb,epsfig,subfigure,color}
\usepackage{array}
\usepackage{booktabs, harvard}

\renewcommand{\le}{\leqslant}
\renewcommand{\ge}{\geqslant}

\newcommand{\be}{\begin{equation}}
\newcommand{\en}{\end{equation}}

\def \p {{\bf{p}}}

\def\p{ {\vec p}}

\def \lam {{\lambda}}

\def \R {{\langle R \rangle}}

\renewcommand{\vec}[1]{\boldsymbol{#1}}
\begin{document}
\numberwithin{equation}{section}

%++++++++++++++++++++++++++++++++++++++++++++++++++=++++++++++++++++++
\title{Methodical Fitting for Mathematical Models \\
of Rubber-like Materials
}
%++++++++++++++++++++++++++++++++++++++++++++++++++++++++++++++++++++

\author{Michel Destrade$^{a}$, Giuseppe Saccomandi$^{a,b}$, Ivonne Sgura$^{c}$ \\[12pt]
 $^a$School of Mathematics, Statistics \& Applied Mathematics,\\
National University of Ireland Galway,
Ireland\\[8pt]
$^b$Dipartimento di Ingegneria, Universit\`a degli Studi di Perugia,\\
Via G. Duranti, Perugia 06125, Italy\\[8pt]
$^c$Dipartimento di Matematica e Fisica ``E. De Giorgi'',\\
Universit\`{a} del Salento, Via per Arnesano, 73100 Lecce,
Italy.}

\date{}

\maketitle

\begin{abstract}

A great variety of models can describe the non-linear response of rubber to uni-axial tension. 
Yet an in-depth understanding of the successive stages of large extension is still lacking. 
We show that the response can be broken down in three steps, which we delineate by relying on a simple formatting of the data, the so-called Mooney transform.
First, the small-to-moderate regime, where \color{black} the polymeric chains unfold easily and \color{black} the Mooney plot is almost linear.
Second, the \color{black} strain-hardening \color{black} regime, where blobs of bundled chains unfold to stiffen the response in correspondence to the ``upturn" of the Mooney plot.
Third, the limiting-chain regime, with a sharp stiffening occurring as the chains extend towards their limit. 
We provide strain-energy functions with terms accounting for each stage, that (i) give an accurate local and then global fitting of the data; (ii) are consistent with weak non-linear elasticity theory; and (iii) can be interpreted in the framework of statistical mechanics.
We apply our method to Treloar's classical experimental data and also to some more recent data. 
Our method not only provides models that describe the experimental data with a very low quantitative relative error, but also shows that the theory of non-linear elasticity is much more robust that seemed at first sight. 
%This modelling is also a fundamental advance of computational solid mechanics, because any reliable numerical simulation of a deformation and/or stress field must rely on the choice of a suitable constitutive relation.

\end{abstract}


\newpage

%%%%%%%%%%%%%%%%%%%%%%%%%

\section{Introduction}

%%%%%%%%%%%%%%%%%%%%%%%%%

The Mechanics of Rubber-like Solids has a  long and prolific history.
Following World War II, a huge research effort was launched to find an explicit strain-energy function able to describe accurately the experimental data obtained from the testing of natural and synthetic rubbers.
However, in spite of decades of intensive work in that area, to this day there is still no effective model able to perform this task in a satisfying and universal way.

This state of affairs is a plain fact, which cannot to be hidden by the countless and seemingly successful models and simulations to be found in the literature.
These simulations may be concretely descriptive but in the end, they apply only to some special phenomena.
From the point of view of physical sciences, constitutive models must be \emph{universal}, not in the sense that  a single model should describe the mechanical behavior of all elastomers, but in the sense that, for a given soft material (e.g. a given sample of natural rubber), a given model should describe its  mechanical response in a satisfactory manner for all deformations fields and all stretch ranges physically attainable.
Here, a \emph{satisfactory} model is defined as a model able to describe the experimental data first of all from a qualitative point of view and then from a quantitative point with acceptable relative errors of prediction with respect to the data.

By scanning all the constitutive models that have been introduced in the literature, we can identify \emph{three fundamental  breakthroughs} over the years.
First, the Mooney-Rivlin strain energy density \cite{mooney}: a purely phenomenological  theory stemming from the early  tremendous effort devoted to rewrite the theory of Continuum Mechanics using the language of Tensor Algebra.
The Mooney-Rivlin model led to the exploration of the non-linear theory of elasticity in deep and unexpected ways, yielded significative classes of non-homogeneous exact solutions and provided a new perspective to the interpretation of experimental data.

The second breakthrough has been the Ogden strain-energy density function \cite{O}: a rational re-elaboration of the Valanis-Landel hypothesis.
For the first time, it became possible to fit accurately theoretical stress-strain curves to experimental data for a variety of deformations and a large range of strains.

The third breakthrough is more complex to describe: it consists in the recent re-elaboration of the ideas underpinning the classical derivation of the neo-Hookean strain-energy based on the basic tools of \emph{statistical mechanics}.
\color{black}
Here there are two possible approaches. 
One is based on micro-mechanical considerations, see for example  \citeasnoun{Pus} for a recent exploration in this direction.
The other is based on molecular considerations, see the detailed paper by \citeasnoun{RuPa02} on the elasticity of polymer networks for a survey.
Below we summarize the micro-mechanical multi-scale approach.

The basic assumption used to derive, from microscopic considerations, the usual models of the mechanical behavior of biological and polymeric networks is the \emph{entropic nature} of their elasticity \cite{Tre}. Because the mechanical response of these materials is due to the deformation of the individual chains or filaments composing the network, there is a strict relationship between the conformations of these macromolecules and the mechanical response of the full macroscopic network.
This situation allows for a simple and direct method to determine the macroscopic strain-energy starting from simple mesoscopic considerations.
Hence the non-linear force-deformation relationship of an ideal chain is easily obtained by considering that the chain's  free energy is purely entropic; then the passage from the single chain model to the full network model
is achieved by means of some \emph{phenomenological} average procedures \cite{Tre}.

Indeed, when we model real macromolecules we rely on several idealized assumptions.
For example, when we assume that there are no interactions between the monomers composing the molecule, we are considering in fact the mathematical model of an \emph{ideal chain}.
It is possible in principle to compute in a careful and detailed way the conformations of such chains in space \cite{Flory}, but these computations are 
cumbersome and usually very complex from a mathematical point of view.
For this reason further \emph{ad hoc} approximations for describing the \emph{end-to-end-distance} of a chain  are introduced \cite{R}.

The most common of these approximations is given by the \emph{Gaussian Chain Model} \cite{Tre}.
The output of this approximation is a linear relationship between the applied force magnitude, $f$,
and the average distance between the chain ends, $\R$, along the direction of the applied force.
This approximation is clearly valid in the small force limit, but the non-linear regime calls for more sophisticated approximations.
Among the various possibilities, two  such non-linear models are very popular:  the \emph{Freely Jointed Chain} model (FJC) for polymers
and  the \emph{Worm-Like Chain} model (WLC)  for stiff biological molecules \cite{R}.

Both the FJC and the WLC models introduce the concept of \emph{contour length}.
The contour length of a polymer chain is its length $R_\text{max}$ at the maximum physically possible extension.
Because  only one configuration can be associated to the maximum extension of the chain in the entropic theory of elasticity, we must have  $f \to\infty$ as $\R \to R_\text{max}$.
This requirement cannot be captured by the Gaussian chain model.
The main difference between the FJC and the WLC models lies in the divergence behaviour of the force as $\R \to R_\text{max}$.  
For the FJC model, we have a first-order singularity: $f \sim (\R - R_\text{max})^{-1}$ whereas for the  WLC model, we have a stronger divergence: $f \sim (\R - R_\text{max})^{-2}$.

\color{black}
In  between the Gaussian Chain model --early stretch regime-- and the FJC or WLC models --late stretch regime-- we find  a stage of \emph{strain hardening} in tension. For instance \citeasnoun{TT} recently summarised this situation for polymers in a good solvent, where the polymeric chains decompose into a succession of independent \emph{blobs}, so named by \citeasnoun{Pincus} who proposed a scaling law $f \sim \R^{3/2}$.
We emphazise that in general the behavior of the polymeric chain is different from the specific one considered by Pincus but that still, a strong strain hardening effect manifests itself when we go from the small deformation regime to the moderate deformation regime.
\color{black}

The aim of this paper is to reconsider the most basic, but still open, problem of rubber-like mechanics: \emph{a meaningful fitting of experimental simple tension data.}
\color{black}
The demand for a model able to describe the entire range of attainable uniaxial tension data is manifest in technological applications of rubber-like materials. 
There, finite element computations play a fundamental role and almost all commercial codes rely on the tangent modulus method.
To be effective, this method needs a model able to capture the \emph{entire} data gleaned from a uniaxial tension test.
\color{black}
We propose to rely on the ideas exposed above for a multiscale approach, by following a rational (and reasonable) procedure in the framework of the non-linear theory of elasticity.

In the next section, we recall the tenets of non-linear elasticity theory, with a focus on modelling the mechanical response of an isotropic incompressible hyperelastic material such as rubber.
We see that the strain energy density $W$ is a function of two variables only, the first two principal invariants of strain and that a reasonable mathematical candidate for $W$ must at least be consistent with fourth-order weak non-linear elasticity. 
For uni-axial data, an historical formatting---the Mooney transform---reveals three distinct regimes of stretch: small-to-moderate, \color{black} strain-hardening\color{black}, limiting-chain, which we delineate rigorously.
We also show that the fitting procedure should rely on minimising the relative errors as opposed to the classical (absolute) residuals, because it provides the most consistent fitting across all regimes and all measures of stress and strain.

\begin{figure}[ht!]
\centering
\includegraphics[width=\textwidth]{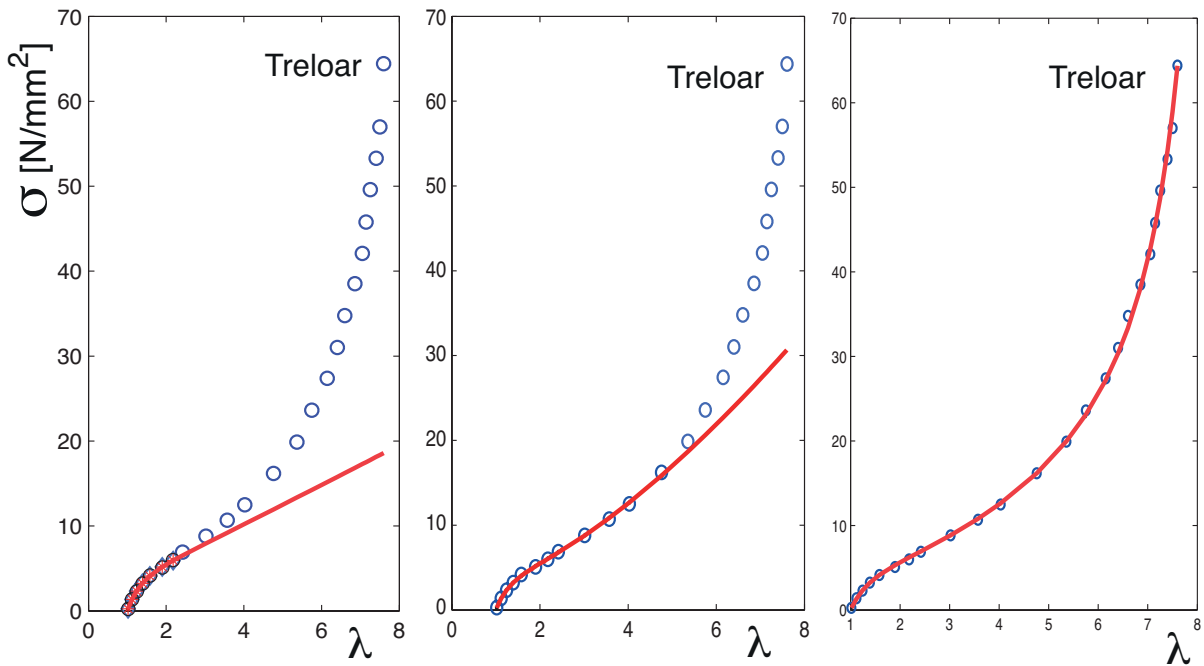}
\caption{
\small
Methodical fitting of the non-linear response of rubber to uni-axial tension (Treloar data).
\color{black} With the Gent-Gent \color{black} strain energy  function, we capture in turn the small-to-moderate strain regime (Section \ref{From small to moderate strains}), the \color{black} strain hardening \color{black} regime (Section \ref{Pincus regime}), and the limiting-chain regime (Section \ref{Limiting chain effect}).
}
\label{fig-recap}
\end{figure}

In Section \ref{Numerical Results} we present two sets of data on the uni-axial extension of rubber: the canonical 1944 data of Treloar (see his book \cite{Tre}) and a more recent set by \citeasnoun{D}. 
Then we proceed to model each regime of stretch in turn.
For the small-to-moderate regime, we show that $W$ must be a function of the two invariants, which rules out the whole class of generalised neo-Hookean materials, for which $W=W(I_1)$ only, including the Yeoh model.
We study three models (Mooney-Rivlin, Gent-Thomas, Carroll) giving excellent fits in that regime.
For the next regime, which corresponds to an upturn in the Mooney plot, we add a \color{black} strain-hardening \color{black} term to those models and again the fit is excellent, now across the resulting wider range of data.
For the range of extreme stretches, we show how to determine the order of singularity, i.e. how to find out whether the rubber stiffens according to the WLC or the FJC model. 
Finally we provide models that fit the data over the entire range of experimental data,
\color{black}
with errors below 4\%.
The conclusion is that the term used to capture the asymptotic limiting-chain behavior also captures the earlier strain-hardening regime, and that the number of fitting parameters can be kept to three only, \color{black} see Figure \ref{fig-recap}.

Section \ref{Conclusion and discussion} is a recapitulation of the results and a reflection on their consequences.

%%%%%%%%%%%%%%%%%%%%%%%

\section{Material modelling}

%%%%%%%%%%%%%%%%%%%%%%%%

Here we present some basic results of the axiomatic theory of continuum mechanics for hyperelastic materials and determine a consistent measure of the goodness of fit for simple tension data.

%%%%%%%%%%%%%%%%%

\subsection{Choice of material parameters}

%%%%%%%%%%%%%%%%%

We consider incompressible isotropic materials which are hyperelastic.
For these it is possible to define a strain-energy density $W=W(I_1, I_2)$, where $I_1=\lambda_1^2+\lambda_2^2+\lambda_3^2$ and  $I_2=\lambda_1^{-2}+\lambda_2^{-2}+\lambda_3^{-2}$ are the principal invariants of the left Cauchy-Green deformation tensor $\vec{B}=\vec{FF}^{T}$, where $\vec{F}$ is the gradient of deformation and the $\lambda_i$ are the principal stretches of the deformation (the square roots of the eigenvalues of $\vec  B$.)
Another representation  of the principal invariants is
\begin{equation}\label{I1I2}
I_1=\text{tr}\left(\vec B\right), \qquad
I_2=\text{tr}\left(\vec B^{-1}\right).
\end{equation}
In passing we note that other sets of invariants exist, but that they are equivalent to our choice for all intents and purposes \cite{sacco15,C}.

This form of the strain-energy density leads to the usual representation formula for the Cauchy stress tensor $\vec T$, due to  \citeasnoun{Rivl96}
\be \label{cauchy}
\vec{T}=-p \vec{I}+2W_1\vec{B}-2W_2\vec{B}^{-1},
\en
where $p$ is the Lagrange multiplier associated with the incompressibility constraint ($\det \vec{F}=1$ at all times) and $W_i=\partial W/\partial I_i$, ($i=1,2$).

Materials  with strain-energy density functions such that $W=W(I_1)$ only are denoted \emph{generalized neo-Hookean materials}.
A prime member of this class is the so-called \emph{neo-Hookean material}
\be \label{neohook}
W_\text{nH}(I_1)=\frac{\mu_0}{2}\left(I_1-3\right),
\en
where the constant $\mu_0>0$ is the infinitesimal shear modulus.

Now, establishing which restrictions should be imposed on the strain-energy function to guarantee reasonable physical behavior is known as Truesdell's \emph{Hauptproblem} of the elasticity theory \cite{Tru}.
In this connection we recall the point of view of \citeasnoun{Ball}:
\begin{quote}
\emph{At the end of the day, perhaps it would have been realized that Hadamard notions of well-posedness are far too restrictive in the nonlinear setting, that non-uniqueness and even non-existence comprise acceptable behavior, and that there are probably no fundamental restrictions on the strain-energy function at all besides those arising from material symmetry and frame-indifference.}
\end{quote}
So that we may not expect much from mathematical requirements.
We must thus resort to \emph{mechanical arguments}.

First of all, we point out that we dealt with frame-indifference and material symmetry when we chose to write $W$ as a function of the principal invariants \eqref{I1I2}, see \citeasnoun{Rivl96}.
Next, we require as a normalization condition that the strain-energy be zero in the reference configuration: $W(3,3)=0$.
Also, for incompressible materials, the requirement that the stress is zero in the reference configuration is always satisfied by a suitable choice of the value of $p$ in that configuration.

The next fundamental requirement is linked to the notion of the \emph{generalized shear modulus function} $\mu$, defined as
\begin{equation} \label{shear-mod1}
\mu = \mu(I_1,I_2) \equiv 2\left(W_1 + W_2 \right).
\end{equation}
In the reference configuration it gives the \emph{infinitesimal shear modulus} $\mu$ as
\begin{equation}
\mu_0 = \mu(3,3).
\label{shear-mod}
\end{equation}
For isotropic incompressible linear materials this is the only significant Lam\'e modulus and compatibility of the non-linear model with the linear theory requires that $\mu_0>0$.

Further, we impose requirements on the  class of models incorporating, at the macroscopic level, information about the contour length of the single chain.
Hence we require that, as the contour length  for the end-to-end distance of the polymeric chain goes to infinity, it is necessary to recover the Gaussian chain model, itself easily connected to the macroscopic neo-Hookean strain energy density \eqref{neohook}.
Hence, suppose that one of the constitutive parameters contained in the strain-energy density, called  $J_m$ (say), is a measure of the model's \emph{limiting-chain extensibility}, the macroscopic analogue of the contour length.
For example in the case of a generalised neo-Hookean material, $J_m$ can be an upper bound for $I_1-3$ which is an average measure of the squared stretch \cite{K}.
In that case we would impose that
\be     \label{001}
\lim_{J_m \rightarrow \infty} W = W_\text{nH}.
\en

A further basic requirement that we may impose on $W$ comes  from the \emph{weakly non-linear theory of elasticity}, where the strain energy is expanded in powers of the strain.
For instance, by choosing the Green-Lagrange tensor $\vec{E}=(\vec{F}^{T}\vec{F}-\vec{I})/2$ as a measure of strain, the most general fourth-order expansion of incompressible elasticity \cite{OO,Hamil04,DeOg10} can be written in the form
\be \label{002}
W= \mu_0 \; \text{tr}(\vec{E}^2)+ \frac{A}{3}\text{tr}(\vec{E}^3)+ D\left(\text{tr}(\vec{E}^2)\right)^2,
\en
where $\mu_0$ is the infinitesimal shear modulus (the only parameter of the linear theory), $A$ is the Landau third-order constant and $D$ is the fourth-order elastic constant.
It is now known that in order to capture fully non-linear effects in solids, the fourth-order theory is the \emph{minimal} model to consider.
For example, in physical acoustics \cite{Norris}, it is necessary to carry out the expansion of $W$ to fourth order to ensure that non-linear effects in shear waves emerge \cite{Hamil04,DeGS11}; in the elastic bending of blocks made of incompressible soft solids, the onset of non-linearity involves third- as well as fourth-order constants \cite{DeGM10}; etc.
In conclusion we will impose a consistent compatibility with the fourth order weakly non-linear theory, instead of  compatibility with linear theory only.

%%%%%%%%%%%%%%%%%%%%%

\subsection{Fitting to uni-axial data}

%%%%%%%%%%%%%%%%%%%%%

From now on we concentrate on the data given by the (idealised) homogeneous deformation of \emph{uni-axial extension} resulting from the application of a uni-directional tension.
We call $\lambda$ the stretch in that direction.

It is a simple matter to compute the first and second invariants of the Cauchy-Green deformation tensors, as
\be
I_1 = \lambda^2 + 2\lambda^{-1}, \qquad I_2 = \lambda^{-2} + 2\lambda,
\en
and the \emph{tensile Cauchy stress component} \cite{O}  $t=T_{11}$, as $t = \lambda \partial W / \partial \lambda$.
In experiments, the current force applied per unit length $f_1$ is recorded, and by dividing it by the reference cross-sectional area of the sample, we arrive at the \emph{engineering tensile stress} $\sigma =\lambda^{-1}t$.
It is given by
\be \label{forse}
\sigma(\lambda)=  \dfrac{\partial W }{\partial \lambda} =  2\left( \lambda
-\lambda^{-2}\right) \left( \dfrac{\partial W }{ \partial I_1} +\lambda^{-1}\dfrac{\partial W }{ \partial I_2} \right).
\en
Then, by plotting the $\lambda-\sigma$ curve for $\lambda \ge 1$, we can perform a curve-fitting exercise for a candidate strain energy density $W$ and access the values of its derivatives.

Alternatively, we could plot the $t-\lambda$ curve to access the derivatives of $W$.

Historically,  \citeasnoun{Rivl96} divided the relationship \eqref{forse} across by $2(\lambda - \lambda^{-2})$ and plotted the curve obtained by having the resulting left handside as the vertical axis coordinate and $\lambda^{-1}$ as the horizontal coordinate.
This modified version of \eqref{forse} is the so-called \emph{Mooney Plot}, given by the transform
\be \label{forsemp}
g(z)=  \dfrac{\partial W }{ \partial I_1} +z \dfrac{\partial W }{ \partial I_2},
\en
where
\be
g(z) \equiv\frac{\sigma(\lambda)}{2\left( \lambda -\lambda^{-2}\right)}, \qquad z =\lambda^{-1}.
\en
In this approach we plot the $z-g$ curve for the range $0<z \le1$ to access the values of the derivatives of $W$.
The practical effect of the Mooney transform is to map the range $1 \le \lambda \le 2$ (where the measurements are the most reliable) over $50\%$ of the $z-$domain.

Now, let us consider the experimental data of stretches and engineering stresses $(\lam_i, \sigma_i)$ for $i=1, \dots, m$ in the \emph{engineering space} $\cal E$, of stretches and Cauchy stresses $(\lam_i, t_i)$ for $i=1, \dots, m$ in the \emph{Cauchy space} $\cal C$,  and the corresponding data in the \emph{Mooney space} $\cal M$, obtained by  the Mooney transform: $(z_i, g_i) :=  \left(\lam_i^{-1}, \sigma_i/[2(\lambda_i-\lambda_i^{-2})]\right)$ for $i=1, \dots, m$, where $m$ is the number of measurements in the uni-axial test.

At this juncture we emphasize that different fitting procedures are possible and that the choice of a given procedure may have an impact on our modeling considerations.

%%%%%%%%%%%%%%%%%%%%%

\subsection{Goodness of fit}

%%%%%%%%%%%%%%%%%%%%%

In general, let  $\vec p =[p_1, \dots, p_n]$ be the set of material parameters involved in the definition of the strain energy $W$, that have to
be identified by \emph{matching}, as best as we can, the data with the desired mathematical model for $W$.
For example, for the neo-Hookean solid \eqref{neohook}  we have $\vec p = [\mu_0]$ and for the fourth-order elasticity model \eqref{002} we have $\vec p = [\mu_0, A, D]$.
To evaluate the best-fit parameters $\vec p$ we solve a Least Squares (LS) problem, which can be linear or non-linear depending on the functional dependence of the strain energy densities $W$ from $\vec p$.

To quantify the goodness of fit, we define on the one hand the following \emph{(absolute) residuals} in the spaces $\cal E$, $\cal C$, and $\cal M$,
\be \label{res}
 \sigma(\lam_i; \vec p) - \sigma_i, \qquad  t(\lam_i; \vec p) - t_i, \qquad g(z_i; \vec p) - g_i, \qquad (i=1, \dots, m),
\en
respectively, and on the other hand, the following \emph{relative residuals},
\begin{equation}
r_i^R(\p)=\frac{\sigma(\lam_i; \p)}{\sigma_i} -1 =\frac{t(\lam_i; \p)}{t_i} -1  = \frac{g(z_i; \p)}{g_i} -1 , \qquad (i=1, \dots, m).
\label{relres}
\end{equation}
The latter equalities, due to the Equations \eqref{forse}-\eqref{forsemp}, show that the \emph{relative residuals are the same in the Engineering, Cauchy and Mooney spaces.}
Moreover, \eqref{relres} shows that the \emph{relative errors} are non-dimensional quantities and can be expressed in percentage, as opposed to the absolute residuals.

Then in the classical LS setting, the strategy  is to minimize either of the following Euclidean norms by varying $\p$,
\begin{equation}\label{LS}
\sum_{i=1}^m \left[ \sigma(\lam_i; \vec p) - \sigma_i \right]^2, \quad   \sum_{i=1}^m \left[ t(\lam_i; \vec p) - t_i \right]^2, \quad  \sum_{i=1}^m \left[g(z_i; \vec p) - g_i\right]^2.
\end{equation}

In this paper we make the choice of \emph{minimising the relative errors} in the two norm, that is
\begin{equation}\label{LSerr}
  \| {\bf r_R}(\p) \|_2^2= \sum_{i=1}^m r_i^R(\p)^2,
\end{equation}
 because it best captures our attempts at optimising the global curve fitting exercise over a large range of stretches. 
It is worth noting that this approach corresponds to a classical weighted LS procedure in the Engineering, Cauchy and Mooney spaces, where the weights are $w_i=1/\sigma_i^2, \ w_i=1/t_i^2, \ w_i=1/g_i^2$, respectively.

We shall leave aside minimisation of classical absolute residuals from now on because we found that it gives large relative errors in the small-to-moderate range of extension, which is not desirable from  experimental and modelling points of view.
Moreover, the best-fit parameters obtained by minimising absolute residuals are not the same when we use the engineering stress data, the Cauchy stress data, or the Mooney plot data, which is a problem as the parameters in a strain energy density should be independent of the choice of stress measure.

To quantify the goodness of our fits, we will record the maximal relative error $\text{err}^*=\| {\bf r_R}(\p^*) \|_\infty$ over the range of interest, defined as follows
\be \label{marerrrel}
\text{err}^* = \max_{i} \left| \dfrac{\sigma(\lam_i; \p^*)}{\sigma_i} - 1\right| = \max_{i} \left| \dfrac{t(\lam_i; \p^*)}{t_i} - 1\right|
=\max_{i}
\left| \dfrac{g(z_i; \p^*)}{g_i}-1\right|.
\en

\begin{figure}[tbp] \centering
\includegraphics[width=0.97\textwidth]{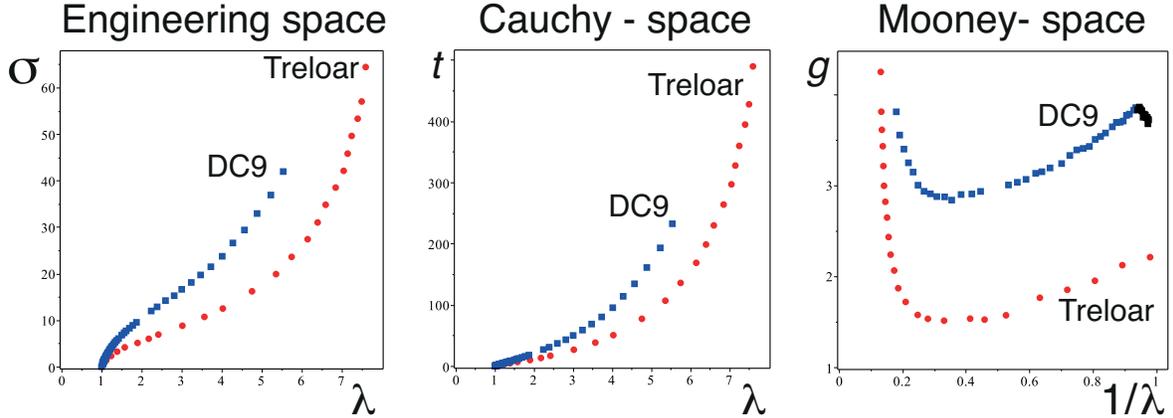}
\caption{
\small
Blue squares: DC9 data; red circles: Treloar's data. 
Left (middle) plot: the tensile engineering (Cauchy) stress $\sigma$ ($t$) in N/mm$^2$ against the stretch $\lambda$.
Right plot: the corresponding $g-z$ Mooney plots.
Three regimes are clearly identified in the $\mathcal M-$space  as we stretch the samples (decreasing $z=1/\lambda$).
Linear decrease: \emph{small-to-moderate regime};  upturn: \emph{\color{black} strain hardening \color{black} regime}; rapid stiffening: \emph{limiting-chain effect regime}.
The first 14 points of the DC9 Mooney plot are greyed out as we must ignore their contribution (see main text).
}\label{F01}
\end{figure}

%%%%%%%%%%%%%%%%%%%%%%%%%%

\section{Numerical Results}
\label{Numerical Results}

%%%%%%%%%%%%%%%%%%%%%%%%%

%=======================

\subsection{Experimental data}

%=======================

To test our models, we will consider two comprehensive sets of data recorded for the uni-axial extension of rubber samples.
The first set is due to \citeasnoun{Tre}, dating back to his canonical 1944 experiments.
For that set we use the original tabulated data, with the engineering stress measured in N/mm$^2$ units, see raw data consisting of 24 points  below.
The second set is more recent and due to \citeasnoun{D}: we will use the data collected in their set labelled `DC9';
it consists of 48 data points.
\begin{multline}
\lambda= [1.02, 1.12, 1.24, 1.39, 1.58, 1.9, 2.18, 2.42, 3.02, 3.57, 4.03, 4.76,\\
 5.36, 5.75, 6.15, 6.4, 6.6, 6.85, 7.05, 7.15, 7.25, 7.4, 7.5, 7.6], 
 \label{lambda-treloar}
\end{multline}
\begin{multline}
\sigma(\lambda)= [0.26, 1.37, 2.3, 3.23, 4.16, 5.1, 6.0, 6.9, 8.8, 10.7, 12.5, 16.2,\\
 19.9, 23.6, 27.4,  31, 34.8, 38.5, 42.1, 45.8, 49.6, 53.3, 57, 64.4].
\end{multline}

Plotting the data in the Mooney space $\mathcal M$ clearly delineates \emph{three regimes of extension} for the elastomers, see Figure \ref{F01}.
Starting from $z = \lambda^{-1}$ at value $1$ (no extension) and decreasing (extension), we see at first a {linear decrease} of $g$ with $z$ for \emph{small-to-moderate deformations}.
Then, around a value for $z$ of $0.3$ (stretch of 230\% or so), an \emph{upturn} occurs as $g$ goes through a minimum.
Finally we enter the \emph{large deformation regime}, where the reduced tensile stress rises sharply and considerably.

Before we move on to model each regime in turn, we note that we had to exclude the first 14 points from the original DC9 data (greyed in the Mooney plot of Figure \ref{F01}), because they prevented the DC9 Mooney plot from having a linear decrease in the small-to-moderate regime. 
For reasons explained later, every soft solid \emph{must} have such a linear decrease.
The reason for the experimental discrepancy presented by the first 14 points is not known, but is probably due to non-optimal accuracy regime for the load-cell in the low load regime and/or the presence of a slight slack of the sample in its starting position.

%=================================

\subsection{From small to moderate strains}
\label{From small to moderate strains}

%=================================

In this section,  we focus on the term in the strain energy that will model the small-to-moderate regime of deformation for the samples.
We identify this region by the \emph{linear part} of the Mooney plot occurring before the upturn, see Figure \ref{F01}.
Hence we perform the curve-fitting exercise on the first $N_0$ data points only, where we determine $N_0$ by conducting a model-dependent sensitivity analysis (to be detailed later).  

\begin{figure}[ht!]
\centering
\includegraphics[width=\textwidth]{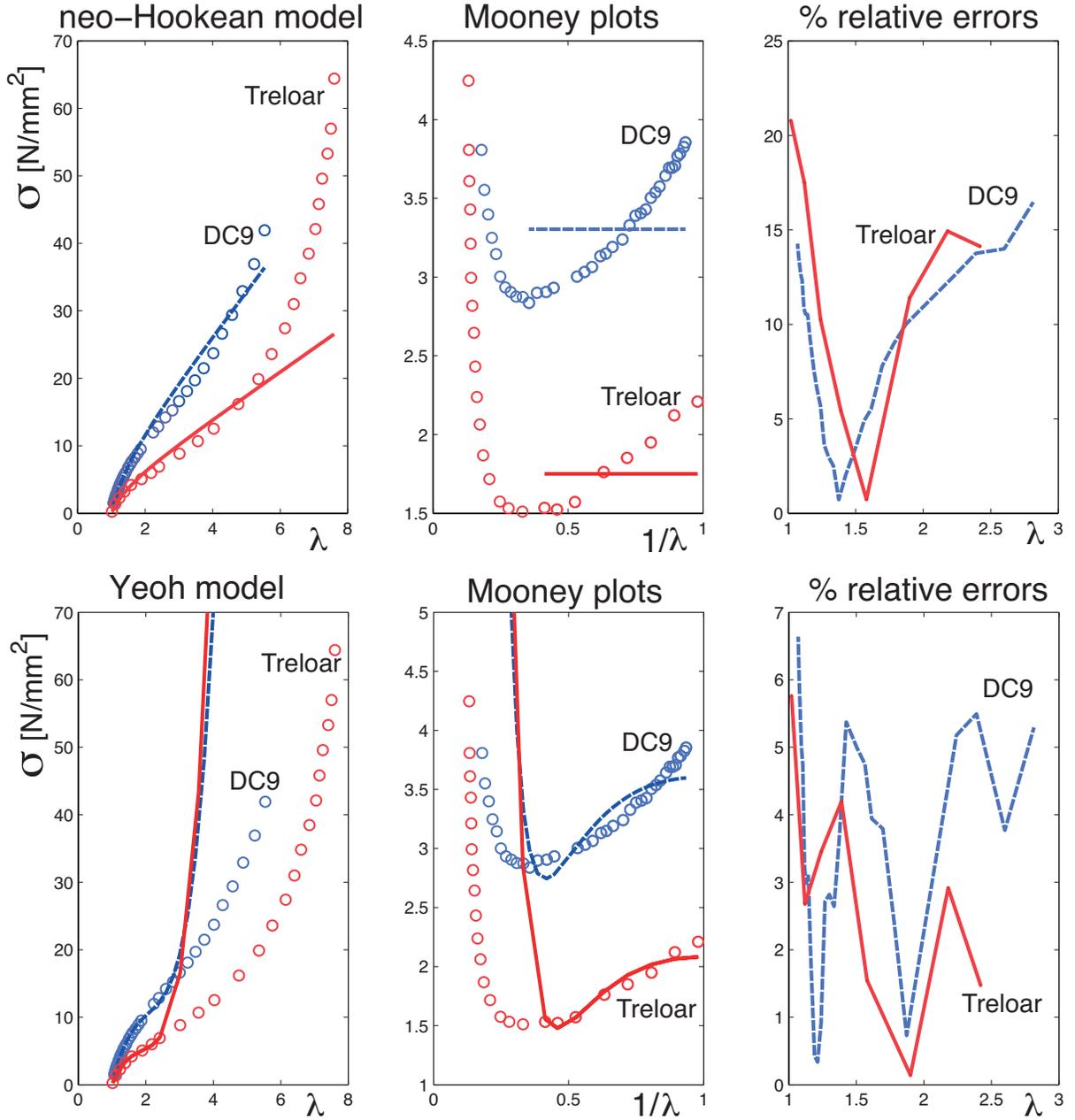}
\caption{
\small
Least-Square fitting of the \emph{neo-Hookean model} (top row) and the \emph{Yeoh model} (bottom row) in the small-to-moderate regime.
For Treloar's data, the best-fit curve is the red full curve and for the DC9 data, the best-fit curve is the blue dashed curve.
Left panels: fitting results in the engineering space on the first $N_0$ data points;
Center panels: the corresponding Mooney plots;
Right panels: corresponding $\%$ relative errors.} \label{FIG1}
\end{figure}

First we see that the neo-Hookean model \eqref{neohook} is incapable of reproducing the simple extension data in any meaningful way.
That is because, for $W_\text{nH}$, the Mooney plot formula \eqref{forsemp} yields
\begin{equation}
g(z) = \mu_0/2,
\end{equation}
giving thus an horizontal line in the $\mathcal M-$space.
The top three panels of Figure \ref{FIG1} illustrate well the resulting unphysical character of the predicted behaviour.
The corresponding relative errors are thus predictably large.

For a polynomial dependence of $W$ on $I_1$, we can consider the Yeoh strain energy density \cite{Yeoh}
\be \label{Yeoh}
W_\text{Yeoh} = c_1(I_1-3) + c_2(I_1-3)^2 + c_3(I_1-3)^3,
\en
where the $c_i$ are constitutive parameters (by \eqref{shear-mod}, the infinitesimal shear modulus of this material is $\mu = 2c_1$).
This model gives of course a better fit in the small-to-moderate region than the neo-Hookean potential, see Figure \ref{FIG1}, lower panels, and Table \ref{table-nH-Yeoh}.

However, as is evident from the formula \eqref{forsemp},
we must include a dependence on the second principal invariant $I_2$ in order to make meaningful progress.

\begin{table}[!htbp]
\centering
\begin{tabular}{l | l l | l l l l }
  material: &  \multicolumn{2}{l |}{neo-Hookean} & \multicolumn{4}{l}{Yeoh}\\
{}   & $\mu_0$   & $\text{err}^*$    & $c_1$   & $c_2$ & $c_3$ & $\text{err}^*$\\
\midrule
Treloar   &  3.7235 & 18.50\%  & 2.1317  & -0.3624 & 0.3624 & \color{black} 3.98\% \\
DC9   &  6.6087 &16.47\% \color{black}  & 3.6048  & -0.2339 & 0.0212 & 6.64\% \\
\bottomrule
\end{tabular}
\caption{
{\small Material parameters for the neo-Hookean and the Yeoh models, obtained by linear curve fitting of Treloar's and DC9 data over the small-to-moderate range (first $N_0=7$ data points for the former and first $N_0=25$ points for the latter). 
The maximal relative error $\text{err}^*$ over the first $N_0$ points is also displayed. 
The corresponding curves and relative errors are shown in Figure \ref{FIG1}. }
}
\label{table-nH-Yeoh}
\end{table}

Hence we will consider in turn the following two classical strain energies: the \emph{Mooney-Rivlin model} \cite{mooney},
\be \label{MR}
W_\text{MR}= \tfrac{1}{2}C_1 (I_1-3)+ \tfrac{1}{2}C_2 (I_2-3),
\en
and the \emph{Gent-Thomas model}  \cite{gent-thomas},
\begin{equation} \label{GT}
W_\text{GT} = \tfrac{1}{2}C_1(I_1-3) + \tfrac{3}{2}C_2 \ln\left(\dfrac{I_2}{3}\right);
\end{equation}
and also the more recent \emph{Carroll model} \cite{C}
\begin{equation}
\label{C}
W_\text{C} = \tfrac{1}{2}C_1(I_1-3) + \sqrt{3}C_2 (\sqrt{I_2}-\sqrt{3}).
\end{equation}
These are two-parameter models, where $C_1$ and $C_2$ are the material constants  to be found from the best-fit procedure.
As noted by \citeasnoun{C}, all three belong to the class of strain energies proposed by \citeasnoun{kling}.
When these strain energies are expanded up to third order in the powers of the Green-Lagrange strain tensor $\vec E$, we find that they all give
\begin{equation}\label{I2-expan}
W = \left(C_1+C_2\right)  \text{tr}(\vec{E}^2) - \tfrac{4}{3}\left(C_1+2C_2\right)  \text{tr}(\vec{E}^3),
\en
giving the following direct correspondence with the  weakly non-linear elasticity expansion \eqref{002}, when stopped at the same order \cite{OO,DeGM10,DeOg10}
\be
\mu_0 = C_1 + C_2, \qquad A = -4(C_1 + 2C_2).
\en

To determine rigorously how many data points we should take for the small-to-moderate region, we performed a sensitivity analysis for all three models.
We found that the maximal relative error is at its lowest value when $N_0=7$ for the Treloar data set  and $N_0=26$ for the DC9 data set, see Figure \ref{fig:sensitivity}.

\begin{figure}[ht!] %
\centering
\includegraphics[width=.9\textwidth]{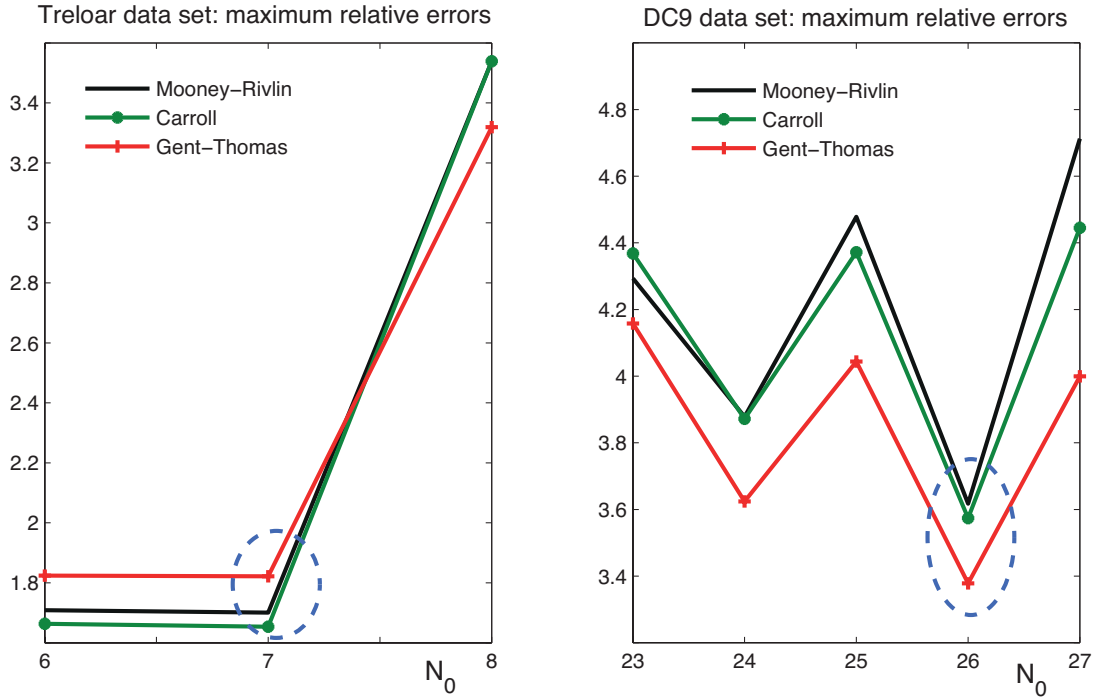}
\caption{
{\small 
Sensitivity analysis to determine the extent of the ``small-to-moderate'' range.
Here the models are fitted in the range $2 \le \lambda \le 3$ and the maximum relative error is recorded in terms of the number of data points considered.}}
\label{fig:sensitivity}
\end{figure}

We see from the resulting best-fit curves of Figure \ref{F3} that, as expected, the $I_2-$dependence is precisely the missing ingredient to obtain excellent agreement in the small-to-moderate regime.
The figure is eloquent on two features.

\begin{figure}[ht!] %
\centering
\includegraphics[width=\textwidth]{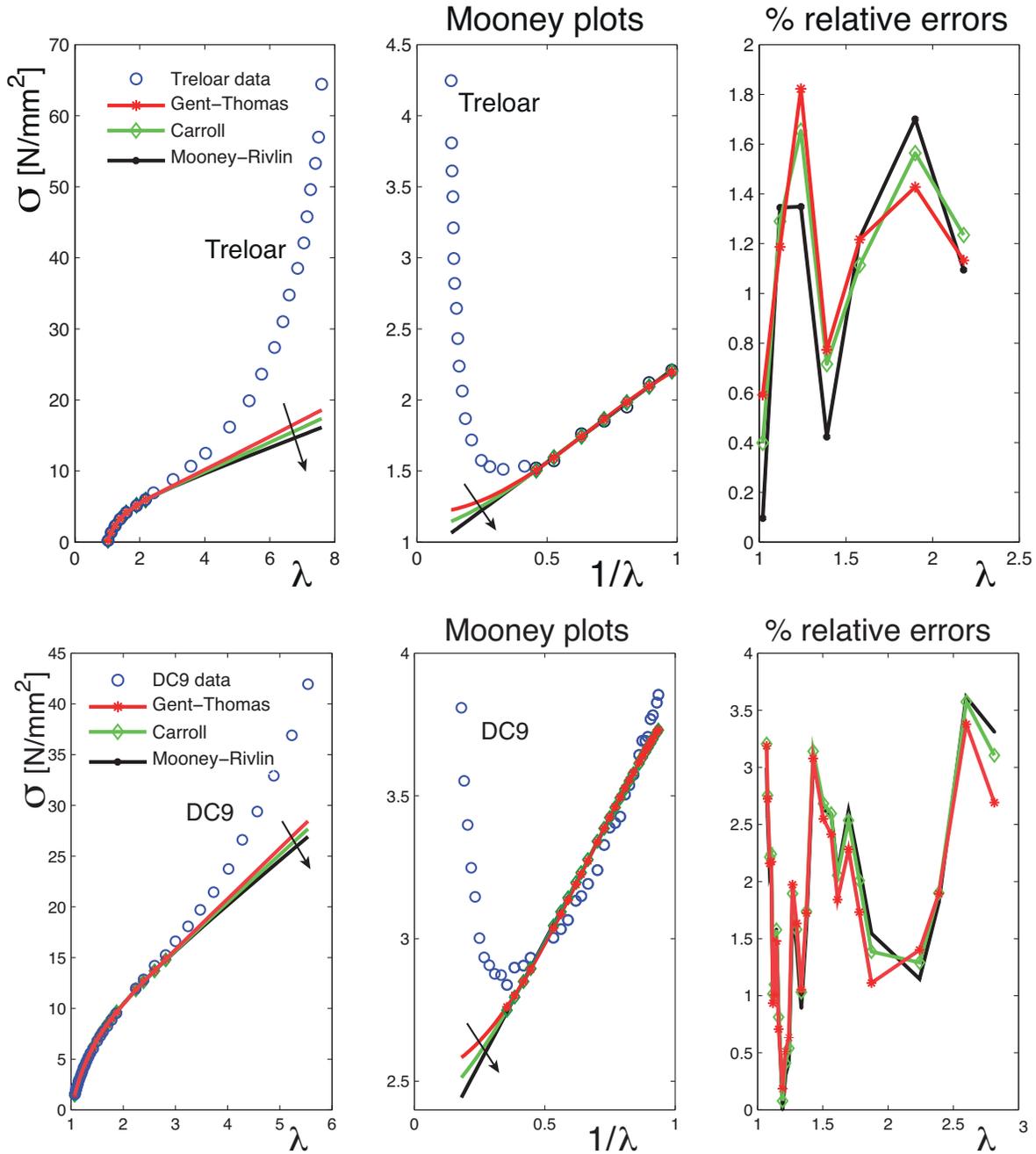}
\caption{
\small
Least-square fitting in the small-to-moderate regime: result of the fitting with $I_2$ corrections by Mooney-Rivlin, Carroll and Gent-Thomas models.
The differences in best-fitting performance between the models are negligible.
Top row: Fitting to Treloar's data;
Bottom row: Fitting to DC9 data.
When the plots are hard to distinguish one from another, an arrow shows the order Gent-Thomas/Carroll/Mooney-Rivlin.}
\label{F3}
\end{figure}

First, the relative errors are dramatically reduced over that regime, to less than $1.8\%$ for Treloar's data and less than $3.6\%$ for the DC9 data, see Table \ref{table-MR-GT-C}.
This improvement is achieved with a set of only two material parameters $\vec p = [C_1, C_2]$ at our disposal, in contrast to the Yeoh model \eqref{Yeoh} which gave much higher errors with a set of three fitting parameters $\vec p =[c_1,c_2,c_3]$.
\begin{table}[!htbp]
\centering
\begin{tabular}{l | l l l l l }
{}   & $C_1$ & $C_2$  & $\text{err}^*$ & $\mu_0$ & $A$    \\
\midrule
  material: &  \multicolumn{5}{l }{Mooney-Rivlin} \\
Treloar   &  1.7725  &  2.7042 & 1.70\% & 4.4767  & -28.724    \\
DC9   & 4.2671  & 3.4329 &  3.62\% & 7.7000 & -44.532 \\
\bottomrule
\toprule
  material: &  \multicolumn{5}{l }{Gent-Thomas} \\
Treloar   & 2.3992 &  2.0348 & 1.82\% & 4.4340 & -25.875 \\
DC9   & 5.0400 & 2.6016 & 3.38\% & 7.6416 &  -40.973 \\
\bottomrule \toprule
  material: &  \multicolumn{5}{l }{Carroll} 
 \\
Treloar   & 2.1580 & 2.2891 & 1.65\% & 4.4471 & -26.945 \\
DC9   & 4.7544 & 2.9006 & 3.57\% & 7.6550 & -42.222 \\
\bottomrule
\end{tabular}
\caption{
\small
Material parameters $C_1$ and $C_2$ (N/mm$^2$) for the Mooney-Rivlin, Gent-Thomas and Carroll models, obtained by linear curve fitting of Treloar's and DC9 data over the small-to-moderate range (first $N_0=7$ data points for the former and first $N_0=26$ points for the latter). 
The maximal relative error $\text{err}^*$ over the first $N_0$ points is also displayed, as well as $\mu_0$ and $A$  (N/mm$^2$), the corresponding  constants of second and third order elasticity, respectively.
The corresponding curves and relative errors are shown in Figure \ref{F3}.
}
\label{table-MR-GT-C}
\end{table}

Second  (as noted earlier by  \citeasnoun{C}), the actual dependence of $W$ on $I_2$ does not matter much, with all three models \eqref{MR}, \eqref{GT}, \eqref{C} performing equally well.
This is not altogether surprising in view of the third-order expansion \eqref{I2-expan} and its universality and relevance to the small-to-moderate range of extension.
In effect, the equivalence of the Mooney-Rivlin strain energy and the third-order expansion shows that every incompressible isotropic hyperelastic solid must present a linear decrease at first in the Mooney space.

In the next section we denote the functional dependence of $W$ on $I_2$  generically by $f(I_2)$ to save space.

%=====================

\subsection{ \color{black} Strain-hardening regime \color{black}}
\label{Pincus regime}

%====================

We now move on to the \color{black}\emph{strain-hardening regime}, \color{black} which is how we called the range of data corresponding to the \emph{upturn} in the Mooney plots, see Figure \ref{F01}.
Thus, here we will perform our curve fitting exercises on the $N_0+N_p$ first data points, where $N_p$ is the number of points required to capture that the Mooney plot has gone through a minimum.
For our two sets of data here, we take $N_p=5$.

Keeping in line with our methodical approach to modeling we now add a \color{black} power-law term \color{black} to the strain energy density of the previous section and consider $W$ in the form
\be \label{MRP}
W = \tfrac{1}{2} C_1 (I_1-3) + C_2f(I_2) + C_3\frac{3^{1-n}}{2n} \left(I_1^n - 3^n\right),
\en
where $C_3>0$  (N/mm$^2$) and $n>0$ (non-dimensional) are constants \cite{oscar}.
As noted by \citeasnoun{C}, as the tensile stretch $\lambda$ increases, the principal force associated with an energy term $I_1^n$ behaves as $\lambda^{2n-1}$
(Note that we could have chosen to add a $I_2^m$ term instead,  in which case the force would behave as $\lambda^{4m-1}$.)
Hence, \color{black} for an ideal (no solvent) polymer, $n=1$ and we recover the Gaussian Chain model; for a polymer in a good solvent, $n=5/4$ and we recover the Pincus correction of the force behaving as $\lambda^{3/2}$.  

Here we wish to model the strain-stiffening as illustrated by the upturn in the Mooney plot. 
As $\lambda$ increases, $I_1^n$ behaves as $\lambda^{2n}$ and $\sigma$ as $\lambda^{2(n-1)}$. 
Hence in the Mooney plot, as $z$ goes towards zero, $g(z)$ behaves as $z^{-2(n-1)}$, indicating that 
\begin{equation}
n>1
\end{equation}
is required for an upturn. 

We point out that the optimisation of the material parameter set $\vec p = [C_1,C_2, C_3,n]$ is a \emph{non-linear} procedure. 
In general, non-linear curve fitting exercises lead to some serious computational problems, the most common being the emergence of multiple minima for the same level of error, see the discussion by \citeasnoun{OgSS04} for rubber models. Also, the choice of an adequate initial guess is a crucial issue in order to avoid false minima \cite{nonlin}. 
To circumvent these problems we propose the following ad hoc procedure.

For definiteness we consider that the optimal value of $n$ should be found in the range $1.0<n<2.5$: the lower bound indicates a departure from the linear fit of the small-to-moderate Mooney plot region and, at twice the Pincus value, the upper bound is a reasonable ceiling to capture the upturn of the Mooney plot prior to the large strain stiffening regime.

We limit the range of data to the $N_0+N_p$ first points, spanning the small-to-moderate regime (linear variation in the Mooney plot) and the strain-hardening regime (upturn in the Mooney plot).
Then we fix $n$ at the beginning of the $1<n< 2.5$ range, at $n=1.01$ say, and perform the fit for the reduced set $\vec p = [C_1,C_2,C_3]$. 
This fit is \emph{linear} and thus removes the risk of multiple minima and the requirement of a good initial guess. 
Then we increase $n$ to span the range and record the corresponding maximum relative errors $\text{err}(n)$.
Finally we keep the optimal value of $n$, corresponding to $\text{err}^* = \min_{1<n< 2.5}\text{err}(n)$. 

For Treloar's data we fix $N_0+N_p=7+5$, and for the DC9 data we take $N_0+N_p=24+5$. 
For the fitting of the former set,  the Gent-Thomas model with a strain-hardening correction gives the lowest error, while for the later set it is the Mooney-Rivlin model with a strain-hardening correction which performs best. 
However, similarly to the previous section, the differences in performance between the three models are negligible, see Table \ref{table-MR-GT-C-P} for a summary.
Overall the fit is excellent over the small-to-moderate and strain-hardening regimes, as attested by the low values of  the relative errors, around or less than 2\%.

\begin{table}[!htbp]
\centering
\begin{tabular}{l | l l l l l l l l}
{}   & $C_1$ & $C_2$  & $C_3$ & $n$ &  $\text{err}^*$  & $\mu_0$ & $A$ & $D$ \\
\midrule
  material &  \multicolumn{5}{l }{Mooney-Rivlin + strain-hardening} \\
Treloar   &   1.7716 & 2.6135 & 0.02725 & 1.6652 & 2.05\%  & 4.412 & -28.10 & 10.83 \\
DC9   & 3.6974  &  3.9934  &  0.0542 &  1.0939 & 1.21\%  & 7.746 & -46.96 & 16.71 \\
\bottomrule
\toprule
  material &  \multicolumn{5}{l }{Gent-Thomas + strain-hardening} \\
Treloar   &  2.0489  & 1.8781  & 0.8911 & 1.6798 & 1.78\% & 4.818 & -26.78 & 8.284 \\
DC9   & 4.6837 & 2.9426 & 0.0434 & 1.0500 & 1.63\% & 7.670 & -42.45 & 13.55 \\
\bottomrule \toprule
  material &  \multicolumn{5}{l }{Carroll + strain-hardening} 
 \\
Treloar & 2.1541 & 2.2080  & 0.02495 & 1.6066 & 1.88\% & 4.387 & -26.38 & 9.612 \\
DC9 & 4.3052 & 3.3351 & 0.0495 & 1.0500 & 1.48\% & 7.6898 & -44.10 &  14.77 \\
\bottomrule
\end{tabular}
\caption{
\small
Material parameters $C_1$, $C_2$, $C_3$  (N/mm$^2$) and $n$ for the Mooney-Rivlin, Gent-Thomas and Carroll models augmented with a \color{black} power-law \color{black} term, obtained by curve fitting of Treloar's and DC9 data over the small-to-moderate range and the \color{black} strain-hardening \color{black} regime (first $N_0+N_p=7+5$ data points for the former and first $N_0+N_p=24+5$ points for the latter). 
The smallest relative error $\text{err}^*$ over the first $N_0+N_p$ points is also displayed, as well as the elastic constants of fourth-order weakly nonlinear elasticity $\mu_0$, $A$ and $D$. 
The corresponding curves and relative errors are shown in Fig. \ref{F4}.}
\label{table-MR-GT-C-P}
\end{table}

In Table \ref{table-MR-GT-C-P} we also report the corresponding constants of weakly non-linear elasticity $\mu_0$, $A$, and $D$. 
They are obtained by expanding $W$ in \eqref{MRP}  up to fourth order in the Green-Lagrange strain tensor. 
We find that now
\be
\mu_0 = C_1 + C_2 + C_3, \quad A = -4(C_1 + 2C_2 + C_3),
\en
while for the fourth-order elastic constant  $D$ in \eqref{002} we have
\be
D= C_1 + 3C_2 +  \tfrac{1}{3}C_3(n+2), \qquad C_1 +  \tfrac{8}{3} C_2 + \tfrac{1}{3}C_3(n+2), \qquad  C_1 + \tfrac{17}{6}C_2 +  \tfrac{1}{3}C_3(n+2),
\en
for the Mooney-Rivlin, Gent-Thomas, and Carroll models with a strain-hardening term, respectively.
We notice that there is no strong continuity for the $C_1$ and $C_2$ parameters from Table 2 to Table 3, reflecting that these parameters are simply the result of a curve fitting exercise. 
The weakly non-linear elasticity constants $\mu_0$ and $A$ however carry consistently from one model to the other.

\color{black}

\begin{center}
\begin{figure}[ht!] %
\includegraphics[width=\textwidth]{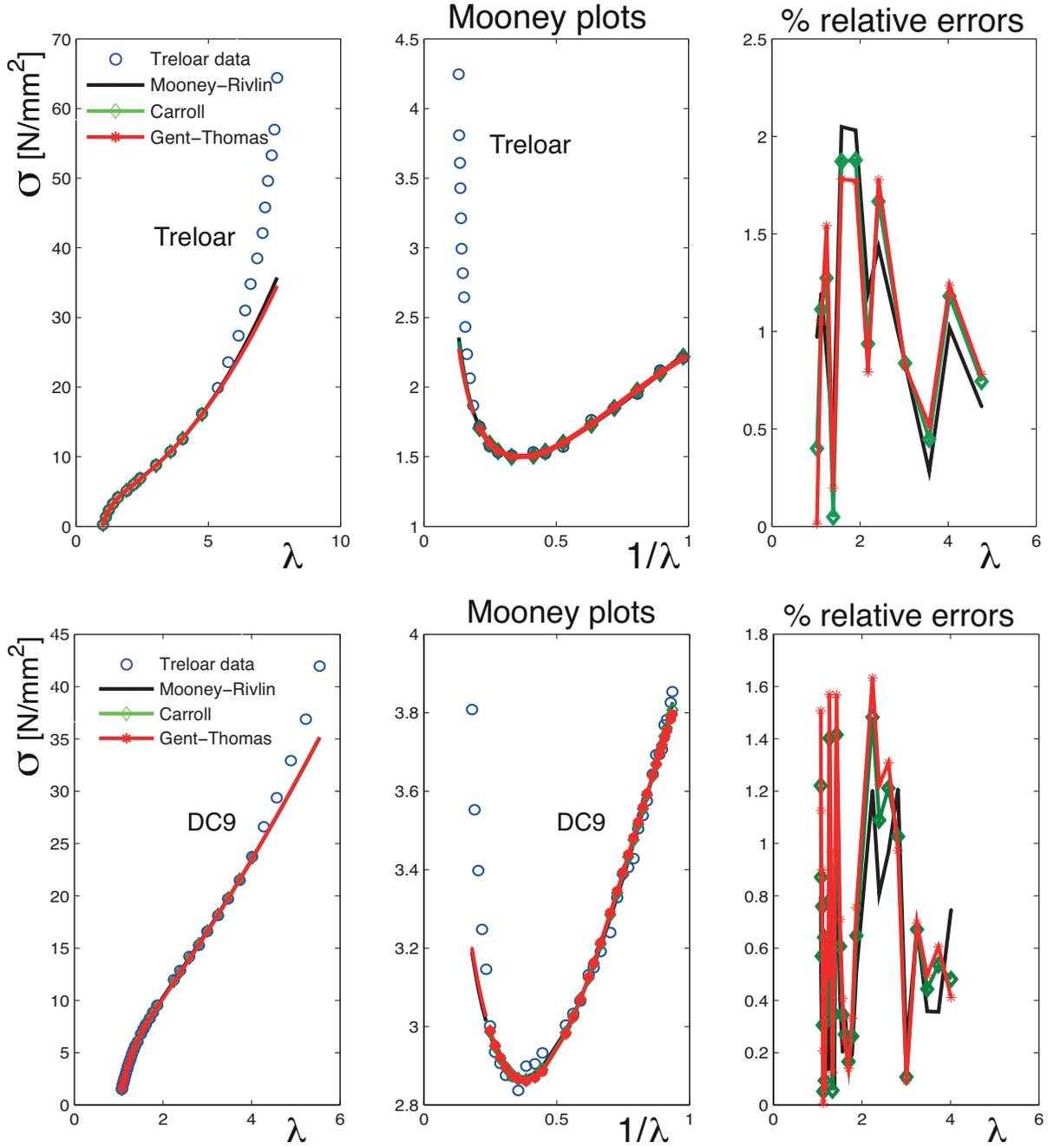}
\caption{
{\small 
Least-square fitting in the small-to-moderate and the \color{black} strain-hardening \color{black} regimes: Mooney-Rivlin, Carroll and Gent-Thomas models with \color{black} power-law \color{black} correction when the first $N_0+N_1$ data points are considered.
Top row: Fitting to Treloar's data (first 7+5 data points); Bottom row: Fitting to DC9 data (first 26+5 points). The `best' fitting results are obtained by the Gent-Thomas model and  by the Mooney-Rivlin model, respectively,  although the differences with the other models are minute and the plots are undistinguishable. 
}
}
\label{F4}
\end{figure}
\end{center}

In this section we showed that \color{black} a strain energy with three terms and four material parameters can \color{black} cover the range of 500\% extension for Treloar's data ($1 \le \lambda \le 6$) and 300\% extension for the DC9 data ($1 \le \lambda \le 4$). 
Is is worth noting that for each model the solution for the optimal set of parameters is unique and the problems are not ill-conditioned.

If we wish to go further and also capture the behaviour of soft matter in a regime of extreme extension ($6 \le \lambda \le 8$ and $4 \le \lambda \le 6$, respectively), we have to recognise that the corresponding stiffening of the curve is associated with a singularity of the fitting function. 
In other words, \emph{limiting-chain effects} come to dominate in the latter regime and impose an asymptotic barrier. 
In the next section we present a method to determine the order of the singularity by examination of the experimental data.

%=================================================

\subsection{Estimate of the order of singularity: FJC vs WLC}

%=================================================

In this section and the next, we focus on the \emph{extreme stretch range}, where the material stiffens rapidly with strain. 
First we consider only the last $N_f$ points of the force-extension data,  and estimate how fast the data curve ``blows up''. 

Assume that in the large stretch range, the tensile force behaves as
\begin{equation} \label{singord}
f = \sigma A_0 \sim \frac{c}{(1-\lambda/\lambda_m)^k}, 
\end{equation}
where $A_0$ is the cross-section area of the sample in the undeformed state, $c$ is a constant, $\lambda = \lambda_m$ indicates the location of the unknown vertical asymptote, and $k$ is the \emph{order of singularity}: $k=1$ for the FJC model and $k=2$ for the WLC model.

In logarithmic coordinates we have
\begin{equation} \label{logcord}
\log(\sigma)=\log(c/A_0)-k\log(1-\lambda/\lambda_m),
\end{equation}
suggesting a linear regression procedure to identify $k$, as follows.
For the last $N_f$ data points, we can calculate $\log(\sigma)$.
We  also know that $\lambda_m$ is greater than $\lambda_f$, the largest stretch reported in the experiments (hence $\lambda_f=7.6$ for the Treloar set, see \eqref{lambda-treloar}, and $\lambda_f = 5.54$ for the DC9 set).
Then we start by fixing $\lambda_m = \lambda_f + 0.001$, say, and plot  $\log(\sigma)$ against  $\log(1-\lambda/\lambda_m)$ to perform a linear fit. We estimate the corresponding maximal relative error and repeat the procedure for a higher value of $\lambda_m$, and so on.

Here for both data sets, we find that the maximal relative error goes through a minimum as $\lambda_m$ increases. 
When it is at its minimal value, we can say that we have identified the actual order of singularity $k^*$ and the actual limiting stretch $\lambda_m^*$.

In Figure \ref{Flog09}, we report these results for the Treloar's and DC9 data, in the upper and lower panels, respectively, where the optimal solutions yielding $k^*$ and $\lambda_m^*$ are highlighted in red.
For Treloar's data, we take $N_f=9$, and find $\lambda_m^*=8.72$ and $k^*=0.962$, with a maximal relative error of $2.1\%$, indicating a \emph{first order singularity} location for large deformations.
For the DC9 rubber data set, we take $N_f = 5$ and find $\lambda_m^*=10.6$, $k^*=2.03$ with a maximal relative error of $0.6\%$, clearly pointing out to a \emph{second-order singularity}.

\begin{center}
\begin{figure}[ht!] %
\centering
 \includegraphics[width=0.8\textwidth]{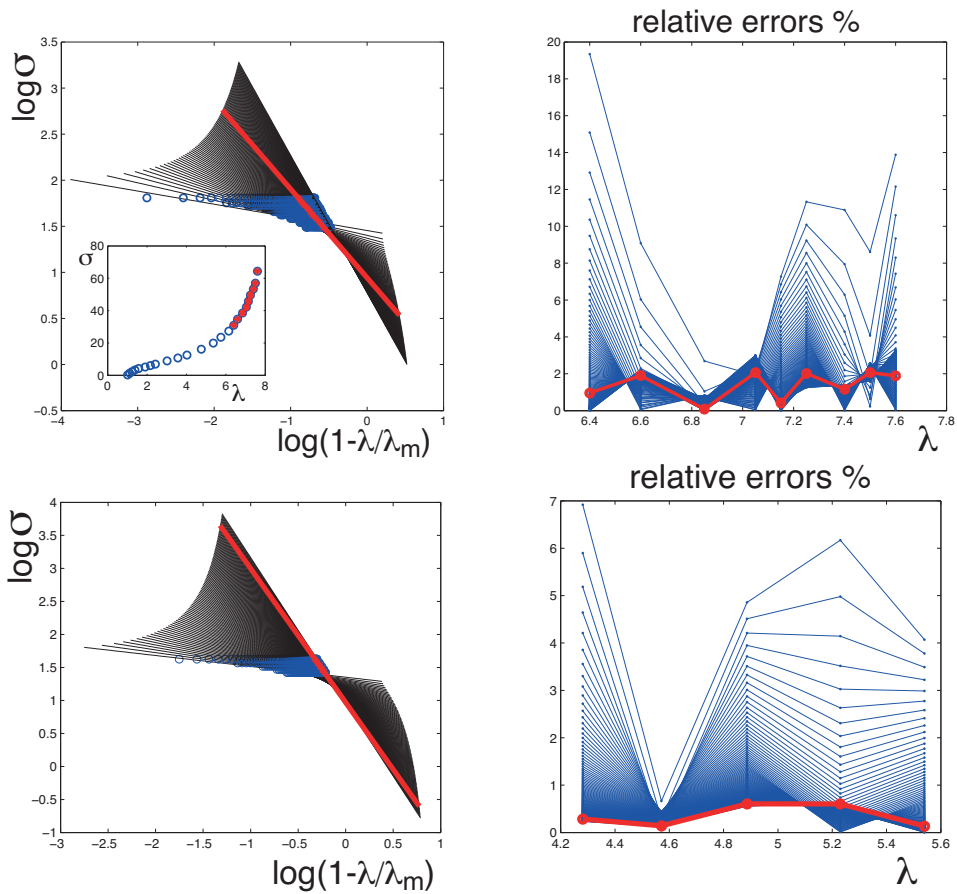}
\caption{
{\small
Estimating the order of singularity $k$ and the maximal stretch of the chain limit $\lambda_m$, using the last 9 points for Treloar's data (top) and the last 5 points for the DC9 data (bottom).  
Left subplots report the linear fittings, right subplots report the corresponding relative errors. The optimal solutions yielding $k^*$ and $\lambda_m^*$ are highlighted by the thick red lines.
}
}
 \label{Flog09} 
\end{figure}
\end{center}

If our goal is to provide a 1-dimensional interpretation of the data at extreme stretches, then our method shows that the rubber studied by Treloar behaves according to the FJC model, while the DC9 rubber extends according to the WLC model.

\color{black} 
The FJC model is easily extended to a 3-dimensional strain energy function, as is often captured by the \citeasnoun{AB} 8-chain model, itself represented well by the phenomenological Gent model \cite{G},
\be \label{Gent}
W=-\tfrac{1}{2}C_1J_m \ln \left(1-\dfrac{I_1-3}{J_m} \right).
\en
Here $C_1>0$ is the initial shear modulus and $J_m>0$ is a stiffening (limiting-chain) parameter.

The WLC model is not easily translated into a 3-dimensional strain energy function, see \citeasnoun{OSS}. 
\citeasnoun{D} proposed to mimic  the WLC behavior with the following phenomenological model
\be \label{WLC}
W=G\left[\frac{I_1}{6}+\beta^{-1}\left(1-\beta \frac{I_1}{3} \right)^{-1} \right],
\en
where $G>0$ and is the ``network infinitesimal shear modulus'' and $\beta$ is a stiffening parameter. 
Note that this model \eqref{WLC} is a special case of the family of models presented by \citeasnoun{Horg1}
(In passing, we also remark that in the absence of the stiffening term, $G$ is the Young modulus, not the shear modulus, see \eqref{shear-mod}.)

Guided by these models, the next section will try to capture the non-linear behavior of the Treloar data and its first-order singularity by invoking the first model above, and the behavior of the DC9 data and its second-order singularity with the second model above.
\color{black}

%%%%%%%%%%%%%%%%%%%%%%%%%%%%%%%%%%

\subsection{Limiting-chain effect}
\label{Limiting chain effect}

%%%%%%%%%%%%%%%%%%%%%%%%%%%%%%%%%%%

In this section, we bring together all our accumulated knowledge to reproduce the non-linear behaviour of the rubber tested by Treloar over the \emph{entire range of experimental stretches}.

\color{black}
The natural summation of our analysis so far is that the whole model to be studied should be given, for the Treloar data, by following form
\be
W= -\tfrac{1}{2}C_1 J_m \ln \left(1-\frac{I_1-3}{J_m}\right) + C_2 f(I_2), 
\label{GGP}
\en
and for the DC9 data, by
\be
W = \frac{(1-\beta^2)}{2(3-2\beta+\beta^2)}C_1\left[I_1 -3 + 6\beta^{-1}\left(1-\beta \frac{I_1}{3} \right)^{-1} - 6\beta^{-1}\left(1-\beta \right)^{-1} \right]+ C_2 f(I_2).
\label{DCP}
\en
(Note that we added constant terms to this expression in order to insure that $W(3,3)=0$.)

\color{black}

Here the main difficulty is to identify the optimal value of the limiting-chain parameters $J_m$ \color{black} or $\beta$, \color{black} because \color{black} their \color{black} presence implies that the fitting procedure is a \emph{non-linear} least-square optimization problem.
A possible consequence is that a set $\vec p = [C_1, C_2, J_m]$ \color{black} or $[C_1, C_2, \beta]$ \color{black} could be wrongly identified as the optimal one, while there exist other minima giving a better or similar error \cite{OgSS04,nonlin}.

To investigate fully the nature of our optimal set, first we solved the non-linear least squares (NLS) problem by applying the {\tt Lsqcurvefit} routine
in Matlab and the {\tt NonlinearFit} routine in Maple, both based on the Levenberg-Marquardt method.
Second, since the models contain only one non-linear term, we were also able to apply the VARPRO method devised by \citeasnoun{Golub}. 
This  `Variable Projection method' algorithm is designed for non-linear LS problems where some of the parameters to be identified appear in linear terms, as here. 
We used a recent Matlab routine \cite{VARPROMatlab}, updated and improved with respect to the original Fortran routine.
\color{black} Third, we used a`semi-linear' approach. 
In that case we identified the values of the constants $C_1$, $C_2$ by a linear fit on the first $N_0+N_1$ data points.
Then we varied the value of the stiffening parameter $J_m$ or $\beta$ and recorded the corresponding maximum relative error.
Then we kept the value of $J_m$ or $\beta$ giving the lowest relative error over the range.

In the end, these three strategies yielded very similar results, with small differences in the values of the constants and with low error, comparable with the experimental error that can be expected from uni-axial tension experiments. 
There is little value in presenting all the results of all three procedures and instead we just present those of the NLS routines, which are the simplest to implement.

For completeness and comparison, we used both models above on both the Treloar and the DC9 data. 
We called them GMR, GG, GC and DCMR, DCG, DCC, corresponding to the Gent \eqref{GGP} or Dobrynin and Carrillo \eqref{DCP} models, respectively, with an $I_2$ dependence of the Mooney-Rivlin, Gent, and Carroll type, in turn.
Notice that the GG model first appeared in \cite{GG} and the DCMR model in \cite{D}.

We also calculated the corresponding constants of second- and third- order elasticity, as
\be
\mu_0 = C_1 + C_2, \qquad A = -4(C_1 + 2C_2), 
\en 
for all models, while the constant of fourth-order elasticity is
\be
D= \tfrac{J_m+1}{J_m}C_1 +3C_2, \quad \tfrac{J_m+1}{J_m} C_1 +  \tfrac{8}{3} C_2, \quad \tfrac{J_m+1}{J_m} C_1  + \tfrac{17}{6}C_2,
\en
for the GMR, GG, and GC models, respectively, 
and
\be
D= \tfrac{13 - 15\beta + 9\beta^2 - 3\beta^3}{3(1-\beta)(3-2\beta+\beta^2)}C_1 +3C_2, \quad \tfrac{13 - 15\beta + 9\beta^2 - 3\beta^3}{3(1-\beta)(3-2\beta+\beta^2)}C_1 +  \tfrac{8}{3} C_2, \quad \tfrac{13 - 15\beta + 9\beta^2 - 3\beta^3}{3(1-\beta)(3-2\beta+\beta^2)}C_1  + \tfrac{17}{6}C_2,
\en
for the DCMR, DCG, and DCC models, respectively.

In Table \ref{table-G-DC-treloar} we collected the results of the curve fitting exercises over the whole range of available data for the Treloar data, and similarly in Table \ref{table-G-DC-DC9} for the DC9 data.

\begin{table}[!htbp]
\centering
\begin{tabular}{l | l l l l l l l }
\midrule
{}   & $C_1$ & $C_2$  & $J_m$ &   $\text{err}^*$  & $\mu_0$ & $A$ & $D$ \\
\midrule
GMR   &   2.1531 & 2.1304 & 74.74 & 5.76\%  & 4.283 & -25.66 & 8.573 \\
GG   &   2.4401 & 1.9511 & 78.33 & 3.38\%  & 4.391 & -25.37 & 7.674 \\
GC    &   2.3319 & 2.0077 & 76.82 & 4.70\%  & 4.340 & -25.39 & 8.051 \\
\bottomrule
\toprule
{}   & $C_1$ & $C_2$  & $\beta$ &   $\text{err}^*$  & $\mu_0$ & $A$ & $D$ \\
\midrule
DCMR   &  1.9542  & 2.2548  & 0.02966 & 5.32\% & 4.209 & -25.85 &  9.631 \\
DCG   & 2.2716 & 2.0134 & 0.02787  & 4.87\% & 4.285 &  -25.19 &  8.699 \\
DCC   & 2.1524 & 2.9426 & 0.02858  & 4.24\% & 4.245 & -25.35 & 8.736 \\
\bottomrule
\end{tabular}
\caption{
\small
\color{black}Material parameters $C_1$, $C_2$  (N/mm$^2$) and $J_m$ or $\beta$ for the Gent (top) or Dobrynin-Carrillo (bottom) version of the Mooney-Rivlin, Gent-Thomas and Carroll models, obtained by curve fitting of Treloar's data over the entire range of available data.
The smallest relative error $\text{err}^*$ is also displayed, as well as the elastic constants of fourth-order weakly nonlinear elasticity $\mu_0$, $A$ and $D$. 
}\color{black}
\label{table-G-DC-treloar}
\end{table}

\begin{table}[!htbp]
\centering
\begin{tabular}{l | l l l l l l l }
\midrule
{}   & $C_1$ & $C_2$  & $J_m$ &   $\text{err}^*$  & $\mu_0$ & $A$ & $D$ \\
\midrule
GMR   &   3.9770 & 3.7662 & 64.16 & 2.70\%  & 7.743 & -46.04 & 15.34 \\
GG   &   4.7472 & 2.9650 & 76.48 & 2.59\%  & 7.712 & -42.71 & 12.72 \\
GC    &   4.4597 & 3.2536 & 70.73 & 2.74\%  & 7.713 & -43.87 & 13.74 \\
\bottomrule
\toprule
{}   & $C_1$ & $C_2$  & $\beta$ &   $\text{err}^*$  & $\mu_0$ & $A$ & $D$ \\
\midrule
DCMR   &  3.8024  & 3.7673  & 0.0327 & 2.69\% & 7.492 & -45.04 &  16.77 \\
DCG   & 4.5773 & 2.9582 & 0.0275  & 2.60\% & 7.456 &  -41.66 &  14.48 \\
DCC   & 4.2062 & 3.2487 & 0.0297  & 2.75\% & 7.455 & -42.81 & 14.83 \\
\bottomrule
\end{tabular}
\caption{
\small
\color{black}Material parameters $C_1$, $C_2$  (N/mm$^2$) and $J_m$ or $\beta$ for the Gent (top) or Dobrynin-Carrillo (bottom) version of the Mooney-Rivlin, Gent-Thomas and Carroll models, obtained by curve fitting of the DC9 data over the entire range of available data.
}\color{black}
\label{table-G-DC-DC9}
\end{table}

For all models, the fit is excellent. 
Hence with  \emph{only three parameters}, we are able to cover the full reported data with an error that compares favourably with the experimental error (recall that Treloar reported values with only three significant digits, see \eqref{lambda-treloar}).
In particular, the  `Gent-Gent model' \cite{OgSS04,GG}
\be
W= -\tfrac{1}{2}C_1 J_m \ln \left(1-\frac{I_1-3}{J_m}\right) + \tfrac{3}{2} C_2 \ln\left(\frac{I_2}{3}\right), 
\label{GG}
\en
gives a maximal relative error of less than 3.4\% \color{black} for the Treloar data and less than 2.6\% for the DC9 data, see plots in Figure \ref{fig-GG}.

In fact we find that the GG model can be used also to capture the small-to-moderate and strain-hardening regimes only (first $N_0+N_p$ data points), so that the model \eqref{MRP} with a power-law term can be advantageously replaced with the Gent-Gent model, see Figure \ref{fig-recap}.

\color{black}

\begin{center}
\begin{figure}[ht!] %
\includegraphics[width=\textwidth]{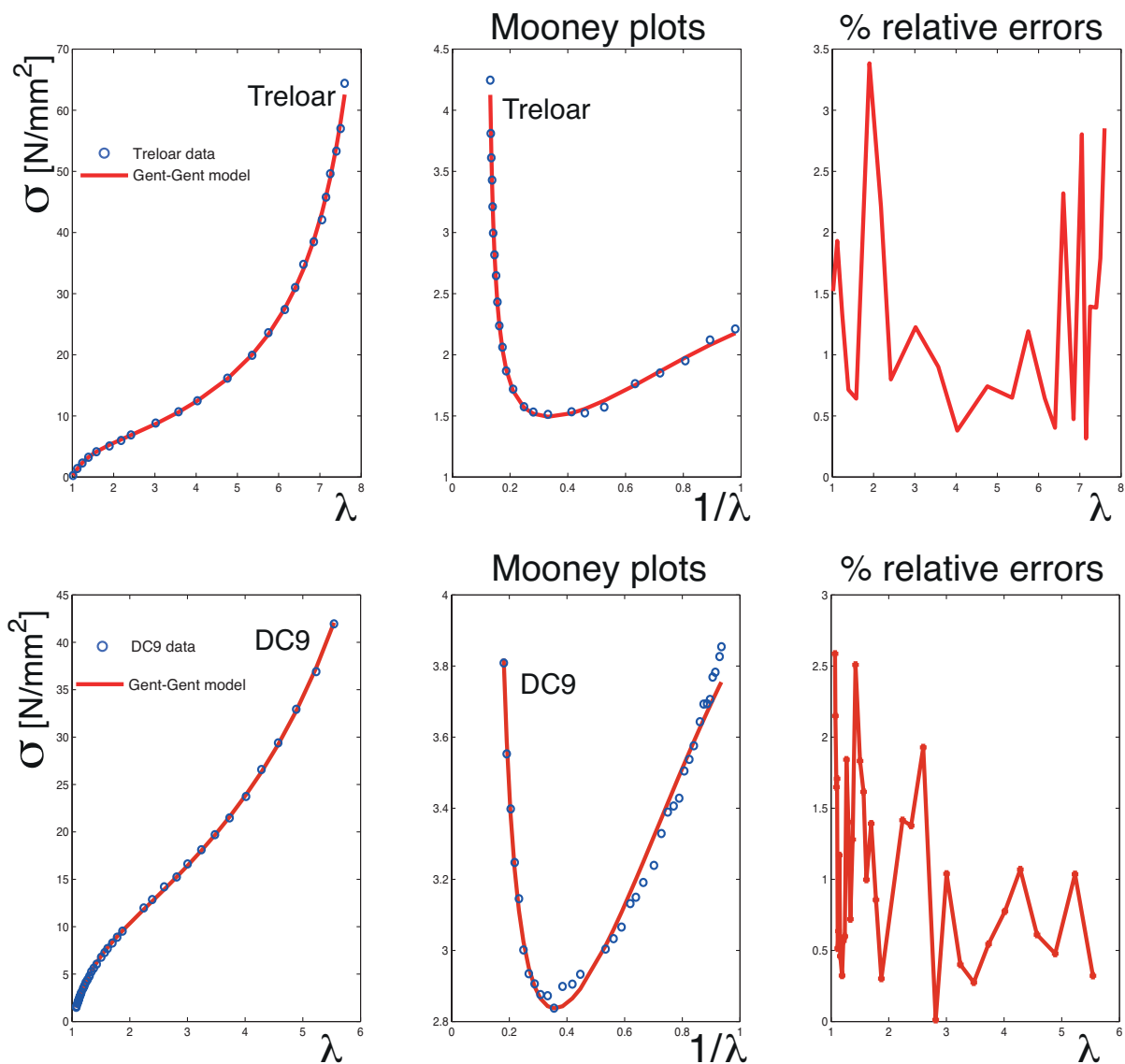}
\caption{
{\small 
Non-linear least-square fitting over the entire range of experimental data with the three-parameter Gent-Gent model.
Top row: Treloar data; Bottom row: DC9 data. 
The corresponding parameters are shown in Table \ref{table-G-DC-treloar}.
}
}
\label{fig-GG}
\end{figure}
\end{center}

%%%%%%%%%%%%%%%%%

\section{Conclusion and discussion}
\label{Conclusion and discussion}

%%%%%%%%%%%%%%%%%%

The fitting of the various constitutive parameters in the mathematical models of rubber-like materials is often considered to be a trivial task, although in fact this is not the case at all.
Indeed more than 10 years ago, \citeasnoun{OgSS04} had already pointed out how the fitting procedure is a very delicate aspect of the modelling procedure. 
Here we went one step further and proposed a possible solution to the main issues. 

We used experimental data already published in literature to show that our method is quite general and does not rely too heavily on the quality of the experimental data to capture   the salient features of the behavior of rubber. 
%We have used the classical Treolar's data because they are a cornerstone in the literature of rubber-like material and then a set of more \emph{modern} data. 
Clearly, it is possible to collect data in a more careful and uniform way for our treatment, for example by fixing a priori the range of extension of interest and by fixing a uniform incremental step in the acquisition of the data for all the samples. 

The starting point  of our method was to delineate three ranges of deformations in the non-linear framework: the small-to-moderate range, the \color{black} strain-hardening \color{black} range (the zone of the upturn point in the Mooney plot), and the region of limiting-chain singularity. 
The aim of the procedure was to find a mathematical model able to fit the data within a reasonable error in part or all of the range. 
The main focus was more on the computational aspect of the fitting procedure than on the corresponding potential experimental issues.
Using a step-by-step method, we eventually came to the conclusion that the theory of non-linear elasticity is a more robust theory than was previously thought.

We saw that minimising the relative errors of \eqref{LSerr}, as opposed to minimising the classical (absolute) residuals of \eqref{LS}, ensures consistency of the fitting procedure across the various measures of stress and strain.

We demonstrated quantitatively that for incompressible materials, the strain energy must depend on both principal invariants $I_1$ and $I_2$. Moreover, we found that any standard linear combination of function of these two invariants fits well the data in the small-to-moderate range. 

Then in order to model the upturn zone, we proposed an additional term to incorporate strain-stiffening effects. 
We also required full compatibility with the weakly non-linear theory of fourth-order elasticity. This compatibility is not a mathematical whim but is dictated by the generality of the response of rubber-like materials to large shear and large bending deformations.  
Again the resulting model provides a robust mathematical behaviour able to capture the data with low relative errors.

For the last step we needed to understand what is going on at very large deformations. 
This range of deformation has been studied in details in the last two decades and two reliable models have emerged: the Worm Like Chain model \cite{OSS} and the Freely Jointed Chain model, itself accurately modelled by the Gent model \cite{G,Pus1}. 
We proposed a practical way to investigate the mathematical behaviour of the vertical asymptote associated with the limiting-chain effects, and thus determine the order of the singularity for a given set of data.
Hence we were able to confirm the insights coming from statistical mechanics computations on ideal macromolecular chains. 

Finally, we showed that the three-parameter Gent-Gent model is capable of fitting the entire deformation range with low relative error for both sets of experimental data studied.
\color{black}
This versatility singles out the Gent-Gent strain energy density as a powerful model for all non-linear aspects of rubber-like behaviour (see also \citeasnoun{OgSS04}, \citeasnoun{GG}, \citeasnoun{MaDe15}), although we note that the other five models investigated also performed well at the curve-fitting tasks. 
\color{black}

By deconstructing, decrypting and then recomposing all the puzzle of mathematical models for rubber-like materials, we provided evidence that the theory of non-linear elasticity can describe accurately uni-axial tension data with a strictly bounded error and a low number of unique parameters. 
The theory of non-linear elasticity is a basic, well-grounded theory of continuum mechanics. 
It is a mandatory passage required to explain the mechanical behaviour of many real-world materials. Therefore it was fundamental to confirm that this theory is not only mathematically well founded but also provides a robust description of experimental data.

Finally, because one of the most important elements of a computer simulation for deformable solid mechanics is the actual model of a material, our findings on the mathematical structure of the mathematical models for rubber-like materials is bound to have non-trivial consequences on their numerical implementation.

%%%%%%%%%%%%%%%%%

\section*{Acknowledgements}

%%%%%%%%%%%%%%%%%%

We are most grateful to Dr Carrillo for providing us with the tabulated version of the data used in his paper \cite{D}.
GS is partially supported by GNFM of Istituto Nazionale di Alta Matematica, Italy.

%++++++++++++++++++++++++++

%%%%%%%%%%%%%%%%%%

\end{document}